\documentclass[showpacs, reprint,
nofootinbib,
amsmath, amssymb,
aps,
]{revtex4-1}

\usepackage{graphicx}
\usepackage{dcolumn}
\usepackage{tabu}
\usepackage{float}
\usepackage{bm}
\usepackage{lipsum}
\usepackage[bookmarks=false]{hyperref}
\usepackage{hyperref}
\hypersetup{pdfstartview={XYZ null null 1.00}}
\hypersetup{
	colorlinks   = true, 
	urlcolor     = blue, 
	linkcolor    = blue, 
	citecolor   = red 
}

\begin{document}

\preprint{APS/123-QED}

\title{Revisiting black holes and wormholes under Weyl transformations}

\author{Fay\c{c}al Hammad}
\email{fhammad@ubishops.ca}
\affiliation{Physics Department \& STAR Research Cluster, Bishop's University,\\
2600 College Street, Sherbrooke, QC, J1M 1Z7, Canada\\
Physics Department, Champlain College-Lennoxville,\\
2580 College Street, Sherbrooke, QC J1M 0C8, Canada}


\begin{abstract}
The behavior of black holes horizon and wormholes under the Weyl conformal transformation is investigated. First, a shorter, but more general, derivation of the Weyl transformation of the simple prescription for detecting horizons and wormholes given recently in the literature for spherically symmetric spacetimes is provided. The derivation allows for a simple and intuitive way to understand why and when horizons and wormholes might arise in the conformal frame even if they were absent in the original frame. Then, the conformal behavior of black holes horizon and wormholes in more general spacetimes, based on more "sophisticated" definitions, is provided. The study shows that black holes and wormholes might always arise in the new frame even if they were absent in the original frame. Moreover, it is shown that some of the definitions found in the literature might be transformed into one another under such transformations. Worked-out examples are given.
\end{abstract}

\pacs {04.50.Kd, 04.70.Bw, 04.20.Jb}
\maketitle
\section{Introduction}\label{SecI}
Both apparent horizons and wormholes have been the subject of much investigation in the field of gravitational physics. While apparent horizons, which arise from the black hole and cosmological solutions of general relativity, are more likely to provide various experimental signatures in the near future (see, e.g., Ref.~\cite{GLS}), their wormholes counterpart might still lie only on the theoretical side for some time. In fact, in contrast to cosmological and black hole spacetimes, which ordinary matter suffices to produce, wormhole spacetimes (more specifically traversable ones) are believed to require matter that violates the energy conditions, and hence dubbed exotic matter \cite{MorrisThorn, VisserBook, Lobo}. The theoretical importance of both apparent horizons and wormholes for our understanding of gravity and spacetime physics is, nevertheless, equal. For this reason, great refinements of their definitions have been proposed in the literature \cite{HaywardTrapping,HVStaticWormhole,HochbergVisser,Hayward,MHC,TIS}.

The other important concept in gravitational physics is the Weyl conformal transformation of spacetime. In contrast to a mere coordinate transformation, under which the physics is left unchanged, a Weyl conformal transformation modifies the geometry of spacetime by modifying lengths and time durations. As such, a Weyl conformal transformation is supposed to modify, not only the formal structure, but the underlying physics as well
\footnote{For a recent study of the subtleties that arise from a Weyl conformal transformation, see Ref.~\cite{BhattacharyaMajhi} concerning its various effects within scalar-tensor theories, and Refs.~\cite{Hammad1, Hammad2} concerning its effect on some of the concepts of mass in general relativity.}
. However, since the debate about whether these transformations do really change the physics has given rise to a large amount of literature in the past without definitively settling the issue (see however Ref.~\cite{ConformalIssue}), we are going to ignore such an issue for now. Our goal in this paper is then to provide a rather more complete investigation than the one already carried out in Ref.~\cite{FPZ} on the relation between the Weyl conformal transformation and wormholes alongside the black holes' horizon counterpart. Our investigation, does in fact cover all the main definitions proposed so far in the literature for wormholes and the black hole horizon.

The outline of the remainder of this paper is as follows. In Sec.~{\ref{SecII}}, we briefly recall the effect of a Weyl transformation on a spherically symmetric spacetime that contains an apparent horizon and/or a wormhole, as derived in Ref.~\cite{FPZ} based on a simple prescription for defining apparent horizons and wormholes. We then provide a shorter, and more general, derivation of the same result in such way that it allows for a more visual and intuitive picture of how and when apparent horizons and wormholes appear in the conformal frame. In Sec.~\ref{SecIII}, we tackle the problem of Weyl transformations of black hole horizons and wormholes in general spacetimes based on more "sophisticated" definitions proposed in the literature. We first recall the main quasilocal definitions proposed, then we examine how these definitions transform under the Weyl rescaling. In Sec.~\ref{SecIV}, we present detailed worked-out examples that illustrate the behavior of each of the definitions under Weyl transformations. We end this paper with a brief conclusion and discussion section.   
\section{A simple "prescription" for spherically symmetric spacetimes}\label{SecII}
It is in order to fully appreciate the subtleties raised by the conformal transformation of black holes' horizon and wormholes in general spacetimes that we are going to devote this entire section to the rather simple, but incomplete, prescription frequently found in the literature for defining apparent horizons and wormholes. 

Any spherically symmetric metric can be represented in a diagonal form with components depending only on the time $t$ and a radial coordinate. One then has the choice of using the physical areal radius $R$, that is the quantity in the metric that multiplies the line element ${\rm{d}}\Theta^2={\rm{d}}\vartheta^2+\sin^2\vartheta {\rm{d}}\varphi^2$ of the unit sphere, as a radial coordinate on which the other components of the metric would depend. The other possibility is to use any other real parameter $r$ as a radial coordinate on which all the components of the metric would depend.

In Ref.~\cite{FPZ}, the authors chose to use the physical areal radius $R$ and made a study of the behavior of horizons and wormhole throats under the Weyl conformal transformation of spacetime given by, 
\begin{equation}\label{Weyl}
\tilde{g}_{\mu\nu}=e^{2\Omega}g_{\mu\nu},
\end{equation}
for any initial spherically symmetric metric $g_{\mu\nu}(x)$ of spacetime and any regular and smooth conformal exponent $\Omega(x)$. Their results showed that a spacetime hosting no horizon and describing a naked singularity in one frame might admit a horizon and describe a wormhole in another frame. Furthermore, the location of the wormhole throats vary from one frame to the other.

The study made in Ref.~\cite{FPZ} relied thus on a general spherically symmetric metric of the form\footnote{Here, and throughout the remainder of this paper we work in units in which $c=1$.},
\begin{equation}\label{SphericallySymmetricMetric}
{\rm{d}}s^2=-A(t,R){\rm{d}}t^2+B(t,R){\rm{d}}R^2+R^2{\rm{d}}\Theta^2,
\end{equation}
where, $A(t,R)$ and $B(t,R)$ are functions of the time coordinate $t$ and the areal radius $R$. Under Weyl's conformal transformation (\ref{Weyl}), this initial line element becomes,
\begin{equation}\label{ConfSphericallySymmetricMetric}
{\rm{d}}\tilde{s}^2=-e^{2\Omega}A(t,R){\rm{d}}t^2+e^{2\Omega}B(t,R){\rm{d}}R^2+\tilde{R}(t)^2{\rm{d}}\Theta^2,
\end{equation}
where, here and in the remainder of this paper, a tilde over a quantity ($\tilde{A}=e^{2\Omega}A, \tilde{B}=e^{2\Omega}B, \tilde{R}=e^{\Omega}R$) or an operator ($\tilde{\nabla}$) refers the object to the conformal frame. 
On the other hand, a simple "prescription" usually taken as a definition for an apparent horizon and/or a wormhole (see Ref.~\cite{FPZ} and the references therein) consists in checking whether the following algebraic equation,
\begin{equation}\label{Prescription}
g^{\mu\nu}\nabla_\mu R\nabla_\mu R=0,
\end{equation}
has a single- or a double-root. One identifies the closed 2-surface corresponding to the single-root with an apparent horizon, while the 2-surface corresponding to the double-root would represent a wormhole throat. While this procedure is incomplete with regard to wormholes and imprecise with regard to apparent horizons, it certainly gives a quick and intuitive way of localizing the radial coordinate of an apparent horizon and/or a wormhole throat. In fact, having a single-root $R_0$ for Eq.~(\ref{Prescription}) just means that the closed sphere of radius $R_0$ would possess the gradient vector $\nabla_\mu R$ that is a null vector. One then expects an apparent horizon. If, on the other hand, Eq.~(\ref{Prescription}) has a double-root $R_*$, one expects to find a wormhole throat as in this case Eq.~(\ref{Prescription}) could be written as $(a^\mu\nabla_\mu R)^2=0=a^\mu\nabla_\mu R$ for some four real parameters $a^\mu$. In fact, the last equality here just means that the areal radius $R$ reaches an extremal value for that specific double-root $R_*$, a feature exhibited by wormhole geometries. Given that the inverse metric $g^{\mu\nu}$ in Eq.~(\ref{Prescription}) is an arbitrary function of $R$ and $t$, that equation might actually have both single- and double-roots simultaneously. This case would then correspond to a coexistence of a wormhole throat and an apparent horizon that might even coincide if the roots do.

Now, while any investigation based on finding single- and double-roots of Eq.~(\ref{Prescription}) renders the task very simple and very intuitive, the procedure is certainly incomplete and imprecise. 

The procedure is incomplete in the sense that for a surface to constitute a wormhole throat it is not sufficient that it be an extremal surface. To be a wormhole throat, it is in fact required to be a minimal surface as well. Furthermore, to be a traversable wormhole, at least by light rays, one imposes on the throat the so-called flare-out condition. The latter condition can be understood in simple terms as a requirement to allow light rays focused by one of the two mouths of the wormhole to come out diverging from the other mouth. 

The procedure is imprecise in the sense that it is not sufficient for a 2-surface to possess a null normal to be a black hole horizon. Black holes' horizons are distinguished from wormhole throats by the fact that they rather constitute the boundary of trapped surfaces. All these additional refinements in the definition of black holes' horizon and wormhole throats will be introduced in more detail in Sec.~\ref{SecIII}. In the present section we only use the requirement (\ref{Prescription}) that we would thus rather call a simple "prescription".    
\subsection{Static case}
Substituting the metric and the areal radius of the initial line element (\ref{SphericallySymmetricMetric}) in the static case into Eq.~(\ref{Prescription}) yields, $1/B=0$. If this equation exhibits a single-root then one can infer that the 2-sphere constitutes an apparent horizon, otherwise one expects a wormhole throat. In the case of multiple double-roots, the wormhole would have a tubular shape as depicted in Fig.~\ref{fig:Figure1} below. In that figure, the wormhole connects two asymptotically flat regions\footnote{It is however not necessary for a wormhole to connect two asymptotically flat regions of spacetime to be called a wormhole as other types of wormholes might exist as well~\cite{VisserBook}.}.
In the case the single-root coincides with the double-root, an apparent horizon would coincide with a wormhole throat or with any other relative minimum surface as depicted in Fig.~\ref{fig:Figure2} below.

After a Weyl transformation, of the form $\Omega=\Omega(R)$ to maintain the static and spherically symmetric character of the spacetime, one easily finds from the new line element (\ref{ConfSphericallySymmetricMetric}) and the prescription (\ref{Prescription}), which transforms into,
\begin{equation}\label{ConfPrescription}
\tilde{g}^{\mu\nu}\tilde{\nabla}_\mu\tilde{R}\tilde{\nabla}_\mu\tilde{R}=0,
\end{equation}
that horizons and/or wormholes in the new conformal frame are conditioned by the solutions to the equation \cite{FPZ},
\begin{equation}\label{SimpleStaticCondition}
\frac{1}{B}(\Omega_{,R}R+1)^2=0.
\end{equation}
To arrive at this equation starting from the metric (\ref{ConfSphericallySymmetricMetric}), however, one has, as is shown in Ref.~\cite{FPZ}, to first turn the line element (\ref{ConfSphericallySymmetricMetric}) of the conformal frame into an expression analogous to the initial line element (\ref{SphericallySymmetricMetric}) but written entirely in terms of the pair $(\tilde{t},\tilde{R})$, instead of the pair $(t,R)$, where $\tilde{t}=\tilde{t}(A,B,R,\Omega)$ is a complicated function of the old quantities. This procedure necessitates in fact a tedious calculation as one obtains a non-diagonal metric first which one then diagonalizes by performing a lengthy coordinate transformation \cite{FPZ}. Furthermore, due to the complex-looking final expression of the new metric components, the authors in Ref.~\cite{FPZ} have focused mainly on the static case. 

Therefore, keeping in mind that condition (\ref{SimpleStaticCondition}) is only valid for time-independent situations, it is easy to see that the possible presence of an apparent horizon and/or wormhole in the old frame ($1/B=0$), is carried over to the new frame with the additional possibility that $\Omega_{,R}R+1=0$ possesses root(s). If the latter is the case, then each of these additional roots is going to be a double-root, giving rise to possibly multiple extremal surfaces, with the minimum one constituting the wormhole throat. An example of the conformal transformation of the wormhole in Fig.~\ref{fig:Figure1} is depicted in Fig.~\ref{fig:Figure3} below, in which the conformal wormhole also has a tubular shape and connects two asymptotically flat regions, but exhibits two relative extrema, two relative minima, and one absolute minimum forming the throat. 

The other important conclusion one draws from condition (\ref{SimpleStaticCondition}) is that in the absence of an apparent horizon in the old frame, \textit{i.e.}, when $1/B=0$ does not admit roots, apparent horizons are forbidden from existing in the conformal frame since this case leads to (\ref{SimpleStaticCondition}) not admitting any single-root either but only double-roots, if any. However, even in the absence of wormholes in the original frame, wormholes could arise in the conformal frame because of the squared parentheses in condition (\ref{SimpleStaticCondition}).

\subsection{Dynamical case}
Let us now examine the dynamical spacetime case by keeping the spherical symmetry assumption. In this case, one uses the full time-dependent metric (\ref{ConfSphericallySymmetricMetric}) after having written it again entirely in terms of the pair $(\tilde{t},\tilde{R})$. In analogy to the static case, one would need only apply the corresponding definition (\ref{ConfPrescription}) to the new metric (\ref{ConfSphericallySymmetricMetric}) thus re-expressed in terms of $(\tilde{t},\tilde{R})$. When doing so, however, the result one finds is $1/\tilde{B}=0$ or, equivalently \cite{FPZ}
\footnote{The expression for $1/\tilde{B}$ that was given in Eq.~(38) of Ref.~\cite{FPZ} was actually more complicated. The form displayed here is what one obtains after doing some rearrangements in that expression.},
\begin{equation}\label{SimpleDynamicCondition}
\frac{1}{B}(\Omega_{,R}R+1)^2-\frac{1}{A}\Omega_{,t}^2R^2=0.
\end{equation}
While this expression is much simpler than expression (38) used in Ref.~\cite{FPZ}, it still does not take explicitly into account the important cases where $R_{,r}=0$ has a root, for which case the areal radius in the old frame reaches an extremum at a given location, and the case where $R_{,t}=0$ has a root, for which the radius' rate of change flips sign. We shall remedy this in the following subsection by making a different choice for the radial coordinate inside the metric.

\subsection{A simpler derivation}\label{SimplerDerivation}
In the remainder of this section, we shall show that, using only fewer and simpler calculations, one can recover not only the same condition about horizons and wormholes in the static case as given by condition (\ref{SimpleStaticCondition}), but also a more explicit and more complete condition, than expression (\ref{SimpleDynamicCondition}) for time-dependent situations. The analysis then becomes simpler, and shows clearly why and when a horizon and/or a wormhole might arise in the new frame even if none exists in the old frame.

Our starting point is again the general spherically symmetric metric, but which we write in terms of the time coordinate $t$ and the radial coordinate $r$ as follows:
\begin{equation}\label{OldMetric}
{\rm{d}}s^2=-A(t,r){\rm{d}}t^2+B(t,r){\rm{d}}r^2+R(t,r)^2{\rm{d}}\Theta^2.
\end{equation}
Under Weyl's conformal transformation (\ref{Weyl}), this metric takes the following form,
\begin{equation}\label{NewMetric}
{\rm{d}}\tilde{s}^2=-e^{2\Omega}A(t,r){\rm{d}}t^2+e^{2\Omega}B(t,r){\rm{d}}r^2+\tilde{R}(t,r)^2{\rm{d}}\Theta^2.
\end{equation}

Let us begin now with the static case. Based on the simple prescription (\ref{Prescription}) for checking the presence of a wormhole and/or an apparent horizon in the initial frame, a double-root would indicate, as explained above, that the areal radius $R$ reaches an extremum somewhere, for which case a wormhole might exist. Actually, an extremum of $R$ is best expressed with respect to the proper radius $l$ such that ${\rm{d}}l=\sqrt{B}{\rm{d}}r$. An extremum of $R$ is then reached in the old frame when ${\rm{d}}R/{\rm{d}}l=0$, that is, when, 
\begin{equation}\label{StaticDR/Dl}
\frac{1}{B}R_{,r}^2=0.
\end{equation}
    This condition is similar to the $1/B=0$ condition found using the metric (\ref{SphericallySymmetricMetric}). The only difference is that our condition now shows {\textit{explicitly}} what a double-root of the defining prescription (\ref{Prescription}) means: it means that $R_{,r}=0$, implying an extremum for $R$. This extra factor comes from the fact that the $1/B$ of Ref.~\cite{FPZ} is, in our case, multiplied by $R_{,r}^2$ because in Ref.~\cite{FPZ} the metric was written in terms of $(t,R)$ instead of $(t,r)$ as is the case here. The effect of this explicit multiplying factor is depicted in Fig.~\ref{fig:Figure1} below. It creates the possibility of having additional extrema besides the true throat, giving a tubular shape for the wormhole. The possibility of having both an apparent horizon ($1/B=0$) and a wormhole ($R_{,r}=0$) is depicted in Fig.~\ref{fig:Figure2}, for which the special case of one relative minimum coinciding with the apparent horizon is illustrated.  
\begin{figure}[H]\centering\includegraphics[angle=0, scale=0.30]{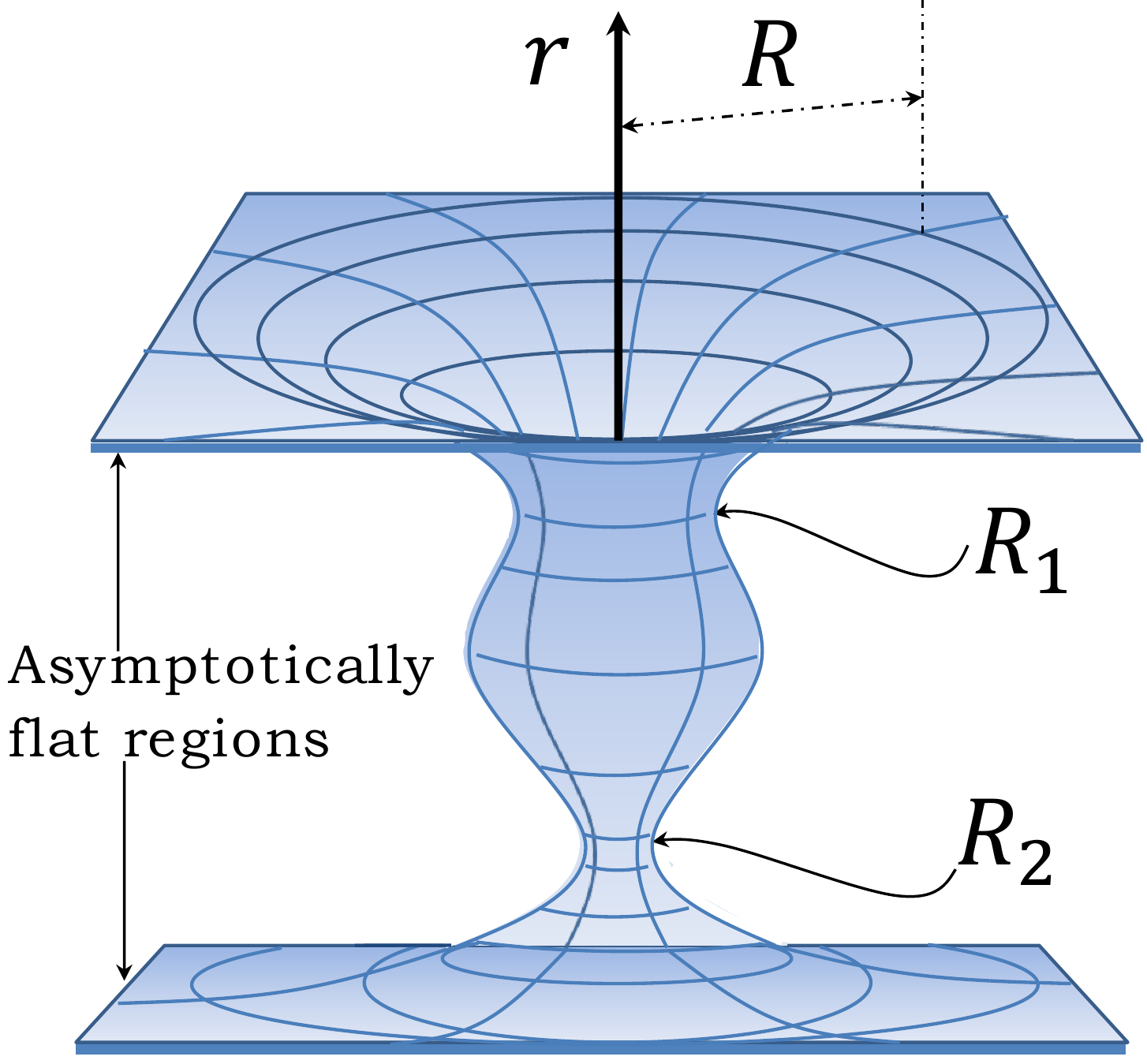}
\caption{Static tubular-shaped wormhole in the original frame with three extremal surfaces, one relative minimum radius $R_1$, one absolute minimum radius $R_2$ making the throat, and one relative maximum in between. In this example, the wormhole connects two asymptotically flat regions.}
\label{fig:Figure1}
\end{figure}
\begin{figure}[H]
\centering\includegraphics[angle=0, scale=0.3]{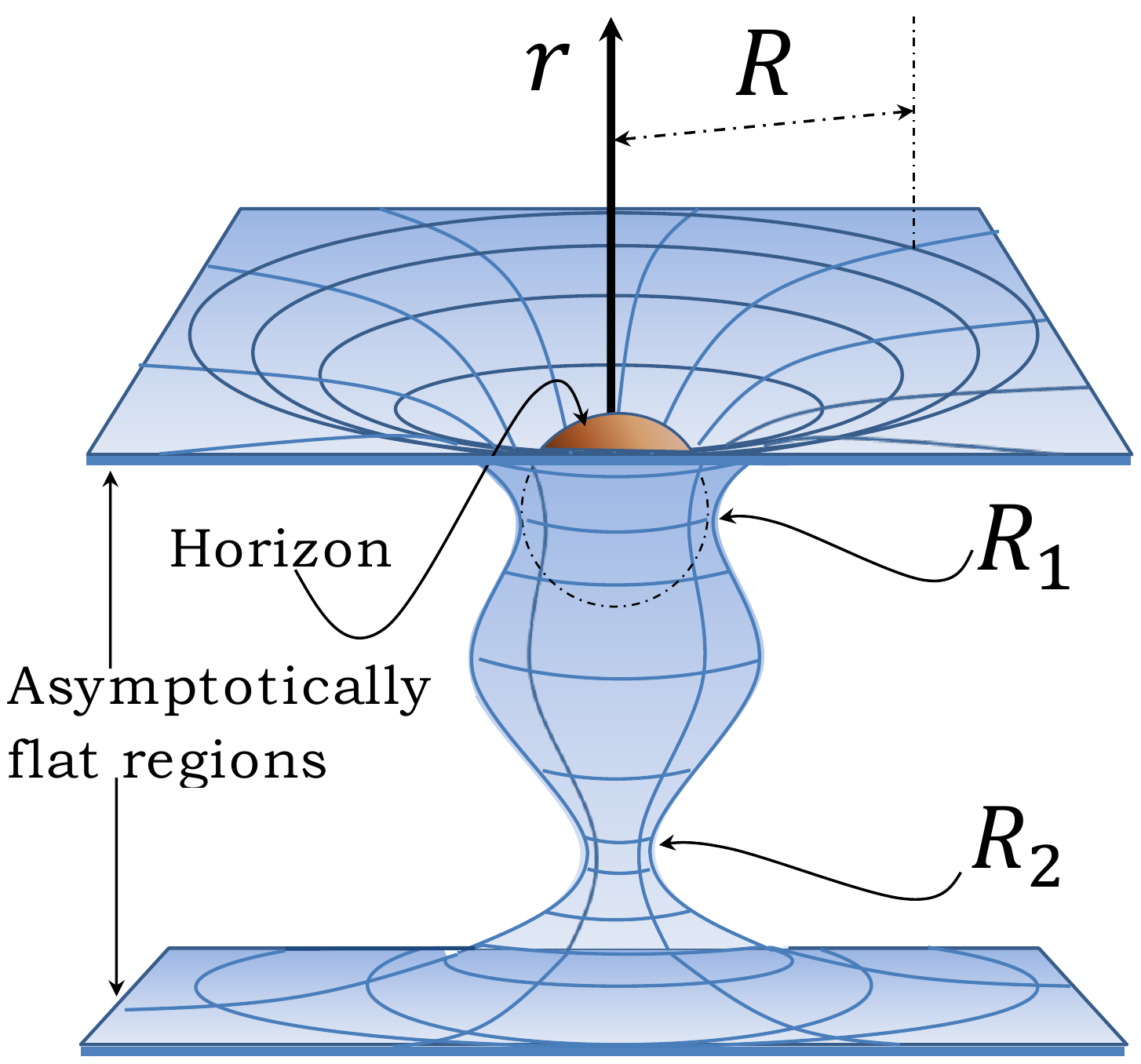}
\caption{Static tubular-shaped wormhole in the original frame with three extremal surfaces, one relative minimum radius $R_1$, one absolute minimum $R_2$ making the throat, and one relative maximum in between. An apparent horizon coincides with one of the minima along the wormhole. In this example, the wormhole connects again two asymptotically flat regions.}
\label{fig:Figure2}
\end{figure}

Analogously, one deduces that an extremum of $\tilde{R}$ is reached in the new frame when ${\rm{d}}\tilde{R}/{\rm{d}}\tilde{l}=0$, that is, when, 
\begin{equation}\label{ConfStaticDR/Dl}
\frac{1}{B}(\Omega_{,r}R+R_{,r})^2=0.
\end{equation}
Again, this condition is similar to condition (\ref{SimpleStaticCondition}) since by using the chain rule, $\Omega_{,r}=\Omega_{,R}R_{,r}$, one recovers the form (\ref{SimpleStaticCondition}) with the additional multiplicative factor $R_{,r}^2$ indicating the origin of any eventual double-root in the original frame that carries over to the new frame.

It must be noted here that both conditions (\ref{StaticDR/Dl}) and (\ref{ConfStaticDR/Dl}) for the original and the conformal frame, respectively, are also recovered by applying the prescriptions (\ref{Prescription}) and (\ref{ConfPrescription}) to the metrics (\ref{OldMetric}) and (\ref{NewMetric}), respectively. This stems from the fact that ${\rm{d}}R/{\rm{d}}l=0$ helps detect both an extremal radius and a null surface. 

Thus, we see that the conclusions drawn using the result (\ref{SimpleStaticCondition}) of Ref.~\cite{FPZ} for the static case still apply here. Namely, because $(\Omega_{,r}R+R_{,r})^2/B=(\Omega_{,R}R+1)^2R_{,r}^2/B$, a horizon in the old frame, obtained when $R_{,r}^2/B=0$ has a single-root (\textit{i.e.}, when $1/B=0$ but $R_{,r}\neq0$) is transformed into an apparent horizon that might coincide with a wormhole throat, obtained when, in addition, $(\Omega_{,R}R+1)^2=0$. But, the absence of an apparent horizon in the old frame ($1/B\neq0$) forbids the appearance of one in the new frame. 

The other important feature is that if there was a wormhole in the old frame, obtained when $R_{,r}^2/B=0$ has a double-root (\textit{i.e.}, $R_{,r}=0$), then such a wormhole always transforms into another wormhole but with more troughs along the tube (\textit{i.e.}, a tubular wormhole) obtained when, in addition, $(\Omega_{,R}R+1)^2=0$. This is illustrated in Fig.~\ref{fig:Figure3} below.
\begin{figure}[H]
\centering\includegraphics[angle=0, scale=0.3]{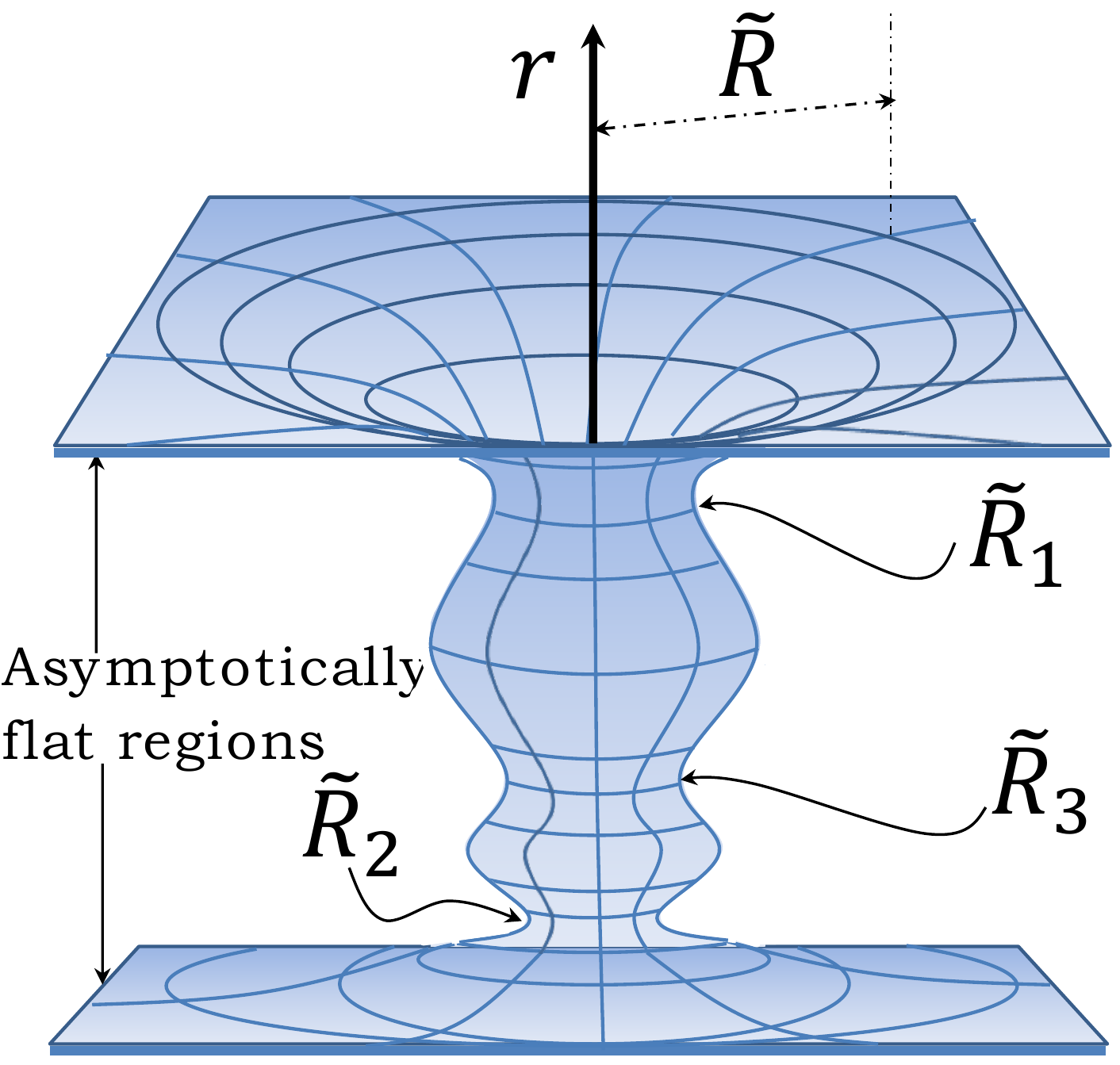}
\caption{Static tubular-shaped wormhole in the conformal frame with five extremal surfaces; two relative minima $\tilde{R}_1$, $\tilde{R}_2$, with one additional trough $\tilde{R}_3$ in between, and one absolute minimum $\tilde{R}_2$ making the throat. In this example, the wormhole connects again two asymptotically flat regions.}
\label{fig:Figure3}
\end{figure}
In case a static apparent horizon coincides with one of the wormhole's static extremal surfaces like in Fig.~\ref{fig:Figure2}, the conformally transformed geometry would always keep this coincidence as depicted in Fig.~\ref{fig:Figure4} below.
\begin{figure}[H]
\centering\includegraphics[angle=0, scale=0.3]{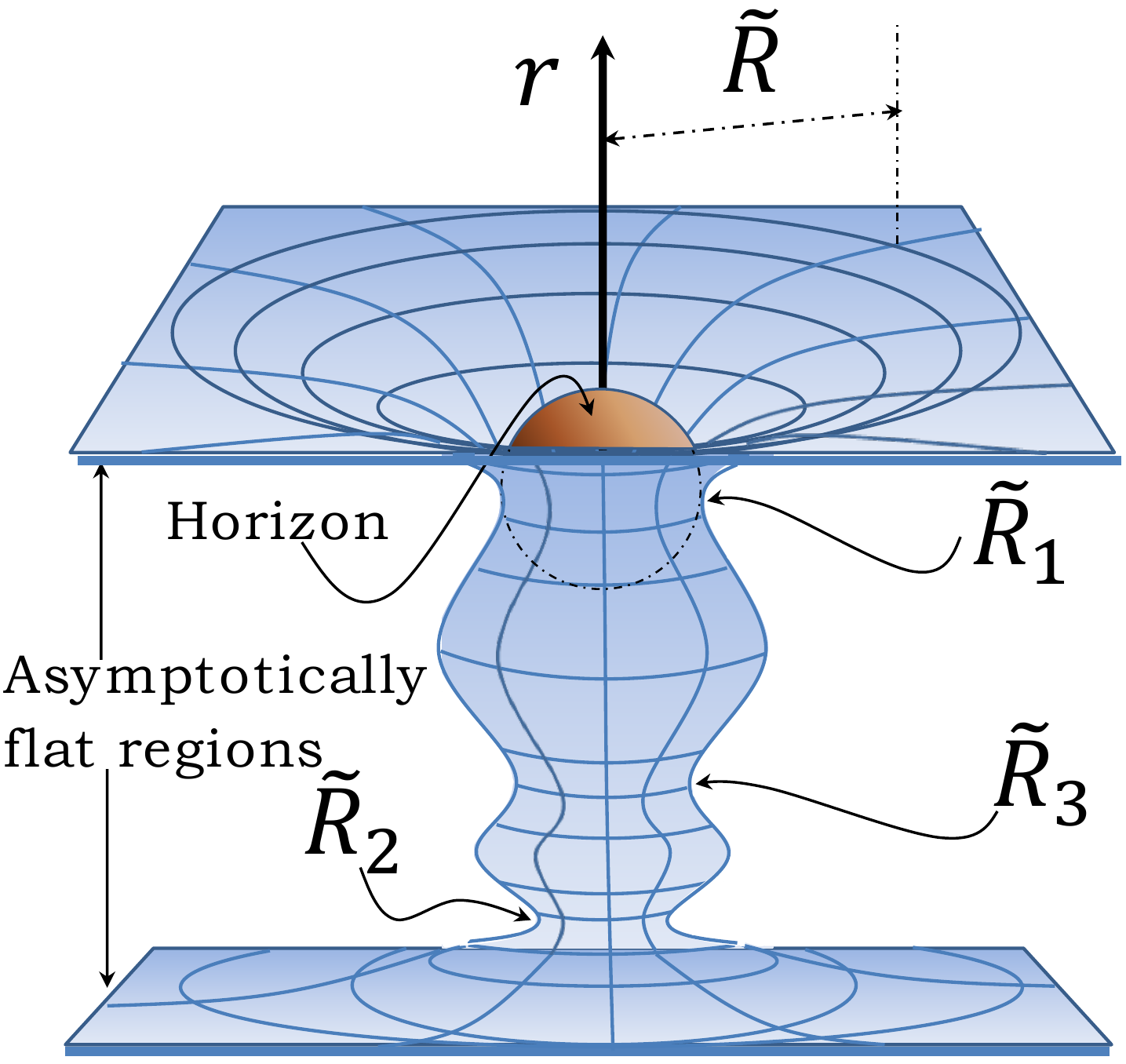}
\caption{Conformally transformed tubular wormhole of Fig.~\ref{fig:Figure3}. The apparent horizon still coincides with one of the extrema; one additional relative minimum and one additional relative maximum appeared along the tube.}
\label{fig:Figure4}
\end{figure}

Let us now turn to the time-dependent case and examine the situation in both frames. When applying the prescription (\ref{Prescription}) to the metric (\ref{OldMetric}) of the initial frame, the condition reads,
\begin{equation}\label{OldCondition}
\frac{1}{B}R_{,r}^2-\frac{1}{A}R_{,t}^2=0.
\end{equation}
Let us provide an interpretation of this condition. For a careful analysis of this condition, however, caution must be made because of the term $R_{,t}^2/A$ that might become undefined when $A(t,r)$ vanishes. Let us therefore start with this special case first.

Whenever $A(t,r)$ vanishes, at one or many instants $t_0$ of time and at the corresponding locations $r_0$, an apparent horizon should be expected as the proper-time lapse suddenly vanishes at those particular instants. At such a particular instant, then, one needs not use (\ref{OldCondition}) to check for apparent horizons and/or wormholes. One needs only check if, in addition, $R_{,r}=0$ has a time-dependent root. If it is the case then one concludes that, besides the time-dependent apparent horizon, a time-dependent wormhole exists in the old frame as well.

Let us now focus on those regions for which $A(t,r)\neq0$. In those regions, condition (\ref{OldCondition}) can fully be used. It is clear that the presence of the time parameter in that equation implies that either the latter is satisfied only at specific instant(s) of time $t_0$ ($t_i, i=1,2,3,...$), at which an apparent horizon and/or a wormhole might come into existence and then disappear, or that the equation is satisfied for all time $t$, for which case an apparent horizon and/or wormhole throat exist at a time-varying location. The coexistence of a wormhole throat with an apparent horizon is obtained when, in addition, one has $R_{,r}=0$. Notice that $A(t,r)\neq0$ ensures that $R_{,r}=0$ would represent a possible genuine wormhole and not just a double-covering of the spacetime outside the horizon as is the case in the Einstein-Rosen bridge, also known as the Schwarzschild wormhole \cite{VisserBook}.

Finally, suppose that equation (\ref{OldCondition}) is not satisfied but still $R_{,r}=0$ might or might not admit roots. This would correspond to the absence of any apparent horizon but with a possibility of the existence of a wormhole.

Let us now go to the new frame and apply the conformal prescription (\ref{ConfPrescription}) on the conformal metric (\ref{NewMetric}). A short and a simple calculation gives,
\begin{equation}\label{NewCondition}
\frac{1}{B}(\Omega_{,r}R+R_{,r})^2-\frac{1}{A}(\Omega_{,t}R+R_{,t})^2=0.
\end{equation}

Suppose again, to start with, that in the old frame the component $A(t,R)$ vanishes somewhere. In this case, one also finds in the new frame that $\tilde{A}(t,r)=e^{2\Omega}A(t,r)$ vanishes at the same time and radial coordinates of the old frame. Therefore, one does not need to use (\ref{NewCondition}) in those regions because an apparent horizon of the old frame would still be present in the new frame, albeit at a different physical location. In addition, however, a time-dependent wormhole might also be present, regardless of the existence of any in the old frame, because the equation $\Omega_{,r}R+R_{,r}=0$, equivalent to $\tilde{R}_{,r}=0$, might admit time-dependent roots which would signal the extremality of the 2-surface, either at all times or at just specific instants of time in the conformal frame.

Suppose now that $A(t,R)\neq0$ everywhere. In this case, condition (\ref{NewCondition}) can be fully used. It is clear here again that the presence of the time parameter in this equation implies that either the latter is satisfied only at specific instant(s) of time $t_0$ ($t_i, i=1,2,3,...$), at which an apparent horizon and/or a wormhole might come into existence in the new frame and then disappear, or that the equation is satisfied for all time $t$, for which case an apparent horizon and/or wormhole throat exist at a time-varying location. The coexistence of a wormhole throat with an apparent horizon in the new frame is obtained when, in addition, one has $\Omega_{,r}R+R_{,r}=0$, or, equivalently, $\tilde{R}_{,r}=0$. Notice again that $\tilde{A}(t,r)\neq0$ ensures that $\tilde{R}_{,r}=0$ would represent a possible genuine wormhole and not just a double-covering of the spacetime outside the horizon.

Now, another very important implication of Eq.~(\ref{NewCondition}) is the following. In contrast to the static case examined above, where the absence of an apparent horizon in the old frame implies its absence in the new frame, in the dynamical case the simple prescription (\ref{ConfPrescription}) based on counting single- and double-roots would allow the creation of an apparent horizon even in the absence of any in the old frame. In fact, according to equation (\ref{NewCondition}), a single-root might be admitted even if none is admitted in the old frame. This possibility arises, however, only in the absence of both wormholes and apparent horizons in the old frame. 

Indeed, suppose that equation (\ref{StaticDR/Dl}) does not admit any solution. In this case neither an apparent horizon nor a wormhole exist in the old frame. On the other hand, this implies that if, in addition $\Omega_{,R}R+1\neq0$, then the only way to satisfy equation (\ref{NewCondition}) is to have $\Omega_{,t}R+R_{,t}\neq0$. Thus, the possibility of having a single-root for the full equation (\ref{NewCondition}) arises, for which case an apparent horizon should then be expected regardless of whether $\tilde{A}(t,r)$ vanishes or not. If now (\ref{StaticDR/Dl}) has only a double-root (no apparent horizon but only a wormhole exists in the old frame), then in the new frame no apparent horizon would arise either because then (\ref{NewCondition}) would also require $(\Omega_{,t}R+R_{,t})^2=0$, which 
entails a double-root, whence a wormhole only.

In summary then, it becomes easy now to see why and when a horizon and/or a wormhole might arise in the new frame as the general scheme can be read off from equation (\ref{NewCondition}). Any apparent horizon that exists in the old frame is automatically carried over to the new frame because $\tilde{A}(t,r)=e^{2\Omega}A(t,r)$: Whenever $A(t,r)$ vanishes the component $\tilde{A}(t,r)$ vanishes also. However, because equation (\ref{NewCondition}) involves a {\textit{difference}} between two squares, it might give rise to a single-root even if the original condition did not have any root, {\textit{i.e.}}, an apparent horizon appears in the new frame even if it is absent in the original frame.
\section{"Sophisticated" Definitions \& General spacetimes}\label{SecIII}
After having studied in detail the simple case of spherically symmetric spacetimes, and acquired some intuition about the behavior of wormhole throats and apparent horizons under the Weyl transformation based on a simple prescription for defining them, we are going to tackle in this section the case of general spacetimes, that are not necessarily static and spherically symmetric. For these cases, no preferred foliation of spacetime is available and one needs to introduce the concept of null geodesics congruences in order to avoid foliation-dependent definitions and ensure generality. Based on this, however, various different definitions have been put forward in the literature \cite{HaywardTrapping,HVStaticWormhole,HochbergVisser,Hayward,MHC,TIS}. We start therefore this section by briefly recalling and motivating these different and more "sophisticated" definitions
\footnote{See Ref.~\cite{BKS} for a nice comparison between the various wormhole definitions examined here.}.
\subsection{Definitions}\label{Definitions}
Before presenting the various definitions, we are going to recall first some properties and formulas for null geodesics and their congruences which will serve us when we perform the Weyl rescaling on those definitions.

Given the two-way orientation of any direction in space, we define a null geodesic by two tangent future null vectors $l_+^\mu$ and $l_-^\mu$; the former for future outgoing light rays and the latter for the future ingoing ones. Because they are null and independent vectors, they are usually chosen to satisfy (the normalization of a null vector being arbitrary though), 
\begin{equation}\label{NullVectors}
l_{\pm\mu} l_\pm^\mu=0 \qquad \text{and} \qquad l_{+\mu} l_-^\mu=-1.
\end{equation}
Here, and henceforth, $\pm$ will mean that the same equation holds when $+$ is replaced by $-$ and {\textit{vice versa}}. Because, they are tangent to geodesics, these vectors automatically satisfy the geodesic equation, which can always be written using an affine parametrization so that the equation takes the form (see, {\textit{e.g.}, Ref.~\cite{Wald}),
\begin{equation}\label{GeodesicEquation}
l_{\pm}^\mu\nabla_\mu l_{\pm}^\nu=0.
\end{equation}
Finally, thanks to these two vectors, any spacetime metric $g_{\mu\nu}$ can actually be decomposed as \cite{Wald},
\begin{equation}\label{TransverseMetric}
g_{\mu\nu}=h_{\mu\nu}-l_{+\mu}l_{-\nu}-l_{-\mu}l_{+\nu},
\end{equation}
where $h_{\mu\nu}$ would be the transverse metric, related to the induced two-metric on the closed 2-surface in which we are interested to check if it constitutes a black hole apparent horizon or a wormhole throat. The two null vectors $l_+^\mu$ and $l_-^\mu$ are normal to the 2-surface and the transverse metric $h_{\mu\nu}$ constitutes a projector onto the 2-surface. One has thus, $l_\pm^\mu h_{\mu\nu}=0$, but also, $h_{\mu\nu}h^{\mu\nu}=2$.

The behavior of a null geodesic congruence is described by the so-called expansion parameter $\theta$ that indicates whether the geodesics of the congruence are diverging from each other ($\theta>0$), converging toward each other ($\theta<0$), or remain parallel to each other ($\theta=0$). This parameter is defined by $\theta_+=h^{\mu\nu}\nabla_\mu l_{+\nu}$ for the outgoing null geodesics and $\theta_-=h^{\mu\nu}\nabla_\mu l_{-\nu}$ for the ingoing null geodesics \cite{Wald}. 
\subsubsection{Black hole horizon}
The definition of a generic black hole horizon that we shall discuss here is that of Hayward \cite{HaywardTrapping} which is defined as the future outer trapping horizon. It is the closure of the hypersurface foliated by marginal surfaces on which the outgoing light rays are instantaneously parallel, while the ingoing light rays are converging outside and inside, or focused. Just inside such a horizon, even the outgoing light rays should be converging. Therefore, the formal definition we use here for the black hole horizon is to have the following three conditions satisfied on the 2-surface $H$ of its horizon \cite{HaywardTrapping},
\begin{equation}\label{BHHorizon}
\theta_+|_H=0,\qquad\theta_-|_H<0,\qquad \partial_-\theta_+|_H<0.
\end{equation}
Here, and in the remainder of this paper, $\partial_\pm$ stands for the derivative with respect to an affine parameter $u^\pm$ along the geodesics whose tangents are $l_\pm^\mu$. Thus, we also have, $\partial_\pm=l_\pm^\mu\partial_\mu$.
\subsubsection{Hochberg-Visser wormhole}
A wormhole, on the other hand, is, as we saw above, an extremal surface. In this section, however, we add the additional more specific requirement that it be a minimal surface. As such, one expects that light rays become focused as they dip into the mouth of the wormhole but expand on the other side as soon as they go past the throat which hosts thus the minimal surface. In other words, a wormhole throat is a marginally anti-trapped surface. Notice that this description does not involve any information about the faraway region outside the throat. Therefore, no requirements such as asymptotic flatness or the global topology of the spacetime hosting the wormhole is involved in such a definition. This makes such a wormhole definition, just as for the case of the black hole horizon, a purely quasilocal geometric definition. 

Thus, the first simple covariant and quasilocal definition of a wormhole that we are going to examine is that of a hypersurface foliated by compact spatial 2-surfaces $S$ on which the following conditions are satisfied, 
\begin{equation}\label{HVWormhole}
\theta_\pm|_S=0 \qquad \text{and} \qquad \partial_\pm\theta_\pm|_S>0.
\end{equation}
This is the definition that has been proposed in Ref.~\cite{HochbergVisser}. A wormhole obeying such a definition we shall therefore call it a Hochberg-Visser wormhole. As shown in detail by the authors there, it turns out that the equation in (\ref{HVWormhole}) expresses the extremality condition of the 2-surface on a null hypersurface, whereas the inequality, which constitutes the equivalent of the "flare-out" condition introduced in Ref.~\cite{MorrisThorn}, expresses the minimality condition of that same 2-surface on that null hypersurface. In view of our later use of it, we are going to recall here the derivation of such a claim. 

Take any compact spatial 2-surface of area $A$ and impose the extremality condition with respect to variations $\delta u^{\pm}$ of the affine parameter $u^{\pm}$ along the outgoing or ingoing null geodesics whose tangent vectors are $l_+^\mu$ and $l_-^\mu$, respectively. The area of the closed 2-surface being given by $A=\int{\rm{d}}^2x\sqrt{h}$, we have,
\begin{equation}\label{SurfaceExtremality}
\delta A=\int{\rm{d}}^2x\sqrt{h}\left(\frac{1}{2}h^{\mu\nu}\frac{{\rm{d}}h_{\mu\nu}}{{\rm{d}}u^{\pm}}\right)\delta u^{\pm}.
\end{equation}
Imposing $\delta A=0$ for all $\delta u^{\pm}$, amounts to imposing the vanishing of the content of the parentheses. On the other hand, the content of the parentheses is just the expansion $\theta_{\pm}$ we introduced above. In fact, since ${\rm{d}}h_{\mu\nu}/{\rm{d}}u^{\pm}=\pounds_{l_{\pm}}h_{\mu\nu}=\pounds_{l_{\pm}}(g_{\mu\nu}+l_{+\mu} l_{-\nu}+l_{-\mu} l_{+\nu})$, where $\pounds_{l_{\pm}}$ stands for the Lie derivative either along the null vector $l_+^\mu$ or the null vector $l_-^\mu$, we have,
\begin{equation}\label{Area/Theta}
\frac{1}{2}h^{\mu\nu}\frac{{\rm{d}}h_{\mu\nu}}{{\rm{d}}u^{\pm}}=h^{\mu\nu}\nabla_\mu l_{\pm\nu}=\theta_{\pm}.
\end{equation}
It should be noted here that in deriving the first equality use has been made of the metric compatibility condition, $\nabla_\rho g_{\mu\nu}=0$, as well as the transversality equation, $h^{\mu\nu}l_{\pm\nu}=0$. Very important for our analysis in Sec.~\ref{SecIV}, though, is the fact that the geodesic equation (\ref{GeodesicEquation}) was not needed to get identity (\ref{Area/Theta}). Thus, we conclude from the latter that $\delta A=0$ is really equivalent to $\theta_{\pm}=0$ on the 2-surface.

As for the minimality of the 2-surface, it is guaranteed if the second variation of the area is positive for all variations $\delta u^{\pm}$. Varying again equality (\ref{SurfaceExtremality}) with respect to $u^\pm$ yields,
\begin{equation}\label{SurfaceMinimality}
\delta^2 A=\int{\rm{d}}^2x\sqrt{h}\left(\theta_\pm^2+\frac{{\rm{d}}\theta_\pm}{{\rm{d}}u^{\pm}}\right)\delta u^{\pm}\delta u^{\pm}.
\end{equation}
Taking into account the fact that extremality already imposes $\theta_\pm=0$, minimality of the surface, corresponding to $\delta^2 A>0$, is then achieved provided that one has, in addition, $\partial_\pm\theta_\pm>0$ 
\footnote{Here we take strict inequalities in order, as already noted in Ref.~\cite{MHC}, to avoid having to identify the Killing horizon of Schwarzschild spacetime with a wormhole throat.}.

\subsubsection{Hayward wormhole}
Another definition put forward in the literature for wormholes is the one given by Hayward in Ref.~\cite{Hayward}. According to this definition, a wormhole throat is a {\textit{timelike}} hypersurface foliated by a non-vanishing minimal spatial 2-surface on a null hypersurface. On such a surface $S$ the conditions, $\theta_\pm=0$ and $\partial_\mp\theta_\pm<0$, must be satisfied. Being a timelike hypersurface, however, one imposes the additional condition, $\zeta^+\partial_+\theta_\pm+\zeta^-\partial_-\theta_\pm=0$, for a timelike vector $\zeta$ such that $\zeta^+\zeta^->0$. When combined with the condition $\partial_\mp\theta_\pm<0$, the timelike requirement of the hypersurface then translates into the condition $\partial_\pm\theta_\pm>0$. Thus, all in all, one requires,
\begin{equation}\label{HWormhole}
\theta_\pm|_S=0, \qquad \partial_\pm\theta_\pm|_S>0, \qquad \partial_\mp\theta_\pm|_S<0.
\end{equation}
In other words, a wormhole throat is a timelike trapping horizon in this definition. It is clear then that this definition includes the Hochberg-Visser wormhole for the case of timelike hypersurfaces \cite{Hayward}. One should keep in mind however that Hochberg-Visser black holes include spacelike hypersurfaces and therefore are not necessarily Hayward wormholes. Therefore, we shall call a wormhole specified by the conditions (\ref{HWormhole}) a Hayward wormhole.

\subsubsection{Maeda-Harada-Carr wormhole}
Both Hochberg-Visser and Hayward wormholes are defined as minimal spatial 2-surfaces on null hypersurfaces and both necessitate the violation of the null energy condition. However, in Ref.~\cite{MHC} a new class of spherically symmetric dynamical wormhole solutions in an accelerating Friedmann background have been found, in which the dominant energy condition is satisfied everywhere. Furthermore, such spacetimes are trapped everywhere but there is no trapping horizon. To include these cosmological wormholes which are asymtotically Friedmann and possess an initial singularity, the authors in Ref.~\cite{MHC} introduced another definition for wormholes. It consists in imposing again extremality and minimality of the 2-surface but with respect to an arbitrary \textit{spacelike} deformation. Thus, the 2-surface is required to be extremal on a spacelike hypersurface. Being suited for cosmological wormholes, the definition introduced in Ref.~\cite{MHC} is only valid for spherically symmetric spacetimes. More precisely, one uses the fact that in the double-null coordinates any spherically symmetric metric can be written in the form,
\begin{equation}\label{MetricForCosmologicalWormholes}
{\rm{d}}s^2=-2e^{2f}{\rm{d}}u{\rm{d}}v+R^2{\rm{d}}\Theta^2,
\end{equation}
where $u$ and $v$ are the null coordinates and $f=f(u,v)$ is a function of these. Then, for any 2-sphere of radius $R=R(u,v)$ to be a wormhole according to this definition consists in imposing the following two conditions on the surface $S$ of such a 2-sphere, expressing, respectively, the extremality and the minimality in the spacelike radial direction $\zeta^\mu$,
\begin{equation}\label{MHCWormhole}
R_{|A}\zeta^A\big|_S=0\qquad \text{and}\qquad R_{|AB}\zeta^A\zeta^B\big|_S>0.
\end{equation}
For convenience, we are using here the same notation as that in Ref.~\cite{MHC}. Namely, the vertical bar with the subscript, $|_A$, stands here for a covariant derivative with respect to the two-metric $g_{AB}$ of the two-dimensional spacetime spanned by the null vectors $\partial_u$ and $\partial_v$. As already noted in Ref.~\cite{MHC}, one should keep in mind that this prescription is dependent on the time slicing. Cosmological wormholes satisfying such conditions we shall call them Maeda-Harada-Carr wormholes.

\subsubsection{Tomikawa-Izumi-Shiromizu wormhole}
Finally, the last definition that has been introduced recently in the literature is that in Ref.~\cite{TIS}. That definition is a kind of mixture between the above three versions. In fact, the authors require the vanishing, not of the expansions themselves, but of the difference between the outgoing and ingoing expansions $\theta_+-\theta_-$. Also, just as in the case of the Maeda-Harada-Carr wormhole, the 2-surface minimality is realized on a spacelike hypersurface. We shall call such wormholes Tomikawa-Izumi-Shiromizu wormholes. Thus, in contrast to the previous definitions this one replaces the extremality and the minimality conditions on the 2-surface $S$ by, respectively,
\begin{equation}\label{TISWormhole}
\theta_+-\theta_-|_S=0\qquad \text{and}\qquad (\partial_+-\partial_-)(\theta_+-\theta_-)|_S>0.
\end{equation}
Here $\partial_\pm$ represent, as before, partial derivatives with respect to the affine parameters $u^\pm$. These can always be chosen as such at least locally. The point behind the inequality in (\ref{TISWormhole}) is that the variation of the expansion is with respect to the spacelike vector $r^\mu=(\partial_+-\partial_-)^\mu$.
We are going now to examine in the remainder of this section the Weyl transformation of black holes' horizon and wormholes based on each of these four definitions.

The important fundamental feature that comes out of these five definitions is that, in contrast to the simple prescription for detecting horizons and wormholes, these more rigorous definitions do not allow a black hole horizon to coincide with a wormhole throat. These definitions make a clear distinction between the two concepts.
\subsection{Transformation of the various definitions}
After a Weyl conformal transformation, not only is the metric $g_{\mu\nu}$ transformed, but, according to the decomposition (\ref{TransverseMetric}), the transverse metric $h_{\mu\nu}$ and the null vectors $l_+^\mu$ and $l_-^\mu$ are as well. As for the transverse metric, it is clear from expression (\ref{TransverseMetric}) that it should transform into $\tilde{h}_{\mu\nu}=e^{2\Omega}h_{\mu\nu}$. For the null vectors, however, there is the usual freedom of rescaling them. In fact, one is free to arbitrary rescale these vectors as long as the first identity in (\ref{NullVectors}) is preserved. To achieve that, one needs only impose for example the following Weyl transformation on the null vectors, $\tilde{l}_\pm^\mu=l_\pm^\mu$ and $\tilde{l}_{\pm\mu}=e^{2\Omega}l_{\pm\mu}$. This option does not preserve the second identity in (\ref{NullVectors}) though. It has been adopted for example in Refs.~\cite{Nielsen}
\footnote{For this choice to work, however, one needs to use, as done in Refs.~\cite{Nielsen}, the generalized version of (\ref{TransverseMetric}) that takes into account the arbitrary normalization of the null vectors. In fact, one needs to use, $g_{\mu\nu}=h_{\mu\nu}+\frac{l_{+\mu}l_{-\nu}}{l_{+\mu}l_-^{\mu}}+\frac{l_{-\mu}l_{+\nu}}{l_{+\mu}l_-^{\mu}}.$}.
However, a more natural choice that we are going to adopt here, which affects both vectors and which preserves the third identity in (\ref{NullVectors}) as well, is the following, $\tilde{l}_\pm^\mu=e^{-\Omega}l_\pm^\mu$, and, as a consequence, we also have, $\tilde{l}_{\pm\mu}=\tilde{g}_{\mu\nu}\tilde{l}_+^\nu=e^{\Omega}l_{\pm\mu}$.

The first consequence of this conformal rescaling of the null vectors is that the geodesic equation (\ref{GeodesicEquation}) is modified into,
\begin{equation}\label{ConfGeodesicEquation}
\tilde{l}_{\pm}^\mu\tilde{\nabla}_\mu \tilde{l}_{\pm}^\nu=\left(\tilde{l}^c\tilde{\nabla}_c\Omega\right)\tilde{l}_\pm^\nu=\left(\frac{{\rm{d}}\Omega}{{\rm{d}}\tilde{u}^\pm}\right)\tilde{l}_\pm^\nu.
\end{equation}
Therefore, after this transformation of the metric the geodesics are no longer affinely parametrized. To restore the affine parametrization, or at least the original parametrization, however, one merely needs to perform a change of parametrization from $\tilde{u}^\pm$ to $u^\pm$ such that,
\begin{equation}\label{ParametrizationChange}
\frac{{\rm{d}}u^{\pm}}{{\rm{d}}\tilde{u}^\pm}=\exp{\int \frac{{\rm{d}}\Omega}{{\rm{d}}\lambda^\pm}{\rm{d}}\lambda^\pm}=e^{\Omega}.
\end{equation}

The next quantities that are affected by this rescaling of the null vectors are the expansions $\theta_\pm$. In fact, a straightforward computation gives the new expansions $\tilde{\theta}_\pm=\tilde{h}^{\mu\nu}\tilde{\nabla}_\mu\tilde{l}_{\pm\nu}$ in the conformal frame in terms of the old expansions $\theta_\pm$ as follows,
\begin{equation}\label{ConfTheta}
\tilde{\theta}_{\pm}=e^{-\Omega}\left(\theta_{\pm}+2l_\pm^\mu\nabla_\mu\Omega\right)=e^{-\Omega}\left(\theta_{\pm}+2\frac{{\rm{d}}\Omega}{{\rm{d}}u^\pm}\right).
\end{equation}
It remains now to check whether this new $\tilde{\theta}_\pm$ is still related to the transverse metric $\tilde{h}^{\mu\nu}$ in the conformal frame as in identity (\ref{Area/Theta}). We have,

\begin{equation}\label{ConfArea/Theta}
\frac{1}{2}\tilde{h}^{\mu\nu}\frac{{\rm{d}}\tilde{h}_{\mu\nu}}{{\rm{d}}\tilde{u}^{\pm}}=\frac{1}{2}h^{\mu\nu}\frac{{\rm{d}}h_{\mu\nu}}{{\rm{d}}\tilde{u}^\pm}+2\frac{\rm{d}\Omega}{{\rm{d}}\tilde{u}^\pm}.
\end{equation}
Changing from the parametrization $\tilde{u}^\pm$ to the affine parametrization $u^\pm$ as defined by the transformation (\ref{ParametrizationChange}) yields the following relation instead,
\begin{equation}\label{NonInvariance}
\frac{1}{2}\tilde{h}^{\mu\nu}\frac{{\rm{d}}\tilde{h}_{\mu\nu}}{{\rm{d}}\tilde{u}^{\pm}}=e^{\Omega}\left(\theta_{\pm}+2\frac{\rm{d}\Omega}{{\rm{d}}u^\pm}\right)=e^{2\Omega}\tilde{\theta}_\pm\neq\tilde{\theta}_\pm.
\end{equation}
This shows that the relation (\ref{Area/Theta}) between the transverse metric, its derivative, and the null congruence's expansion is not invariant under Weyl's conformal transformation. Notice that even if we choose the scaling $\tilde{l}_{\pm\mu}=e^{2\Omega}l_{\pm\mu}$ for the null vectors, the invariance of the relation would not be achieved. Only the "unnatural" rescaling $\tilde{l}_{\pm\mu}=e^{3\Omega}l_{\pm\mu}$ would keep this relation invariant.
However, because the deviation in equality (\ref{NonInvariance}) consists only of the positive and non-vanishing multiplicative factor $e^{2\Omega}$, the extremality of the closed 2-surface is still guaranteed -- as in the original frame -- by the vanishing of $\tilde{\theta}_\pm$ while the minimality of that surface would also be guaranteed by ${\rm{d}}\tilde{\theta}_\pm/{\rm{d}}\tilde{u}^\pm>0.$ 

Let us now go over all our previous definitions of the black hole horizon and wormholes and examine their behavior under Weyl's conformal transformation. 
Starting with the definition (\ref{BHHorizon}) for a generic black hole horizon, we easily find by using the result (\ref{ConfTheta}) that for a black hole horizon to exist in the new frame, one needs all three of the following conditions to be satisfied,
\begin{equation}
\begin{gathered}\label{ConfBHHorizon}
\theta_++2\partial_+\Omega\big|_H=0,\qquad
\theta_-+2\partial_-\Omega\big|_H<0,\\
\partial_-\theta_++2\partial_-\partial_+\Omega\big|_H<0.
\end{gathered}
\end{equation}
It is clear that it is always possible for the first equality to hold, and therefore for a black hole horizon to arise, in the new frame for a given conformal exponent $\Omega$. The two remaining inequalities in (\ref{ConfBHHorizon}) just add constraints on the "degree" of negativity of the original ingoing expansion and the "degree" of convergence of outgoing light rays inside the original horizon in order to have one in the new frame. Moreover, from the first equality we deduce that a black hole horizon might still arise even if none exists in the old frame ({\textit{i.e.}}, $\theta_+\neq0$) provided only that the conformal exponent $\Omega$ does vary with $u^+$. That is, $l_+^\mu\partial_\mu\Omega\neq0$. This clearly departs from what the simple prescription (\ref{Prescription}) allows. In fact, as we saw in the previous section black holes horizons might arise in the new frame when none exists in the old frame {\textit{only}} for time-dependent transformations. This being forbidden when the conformal frame remains static. Keeping in mind that the prescription (\ref{Prescription}) is only valid for spherically symmetric spacetimes, however, we shall see in the worked out examples below that the rigorous definition (\ref{BHHorizon}) also forbids it when conformally transforming into such static spacetimes.

Let us now move on to the Hochberg-Visser wormhole. Using the definition (\ref{HVWormhole}) and the result (\ref{ConfTheta}) we easily verify that to have such a wormhole in the conformal frame we need the following two conditions to be satisfied on the 2-surface that is supposed to represent the wormhole's throat, 
\begin{equation}\label{ConfHVWormhole}
\theta_{\pm}+2\partial_\pm\Omega\big|_S=0\quad \text{and} \quad \partial_\pm\theta_{\pm}+2\partial_\pm\partial_\pm\Omega\big|_S>0.
\end{equation}
The interpretation of these conditions is again straightforward and is as follows. To have a wormhole in the new frame if one already exists in the old frame ({\textit{i.e.}}, $\theta_\pm=0$) the equality in (\ref{ConfHVWormhole}) requires the conformal factor to be independent of the parameter $u^\pm$, that is, $l_+^\mu\partial_\mu\Omega=0$. The inequality, on the other hand, just constrains the degree of the flare-out condition in the original frame. Moreover, we see that if no Hochberg-Visser wormhole exists in the old frame, {\textit{i.e.}}, none of the conditions (\ref{HVWormhole}) are satisfied, such a wormhole might still appear in the new frame, provided that $\Omega$ is adequately chosen.   

Let us now examine Hayward's wormhole definition. First, recall that Hayward's wormhole should still be a timelike hypersurface in the conformal frame. Therefore, the expansion in the new frame needs to satisfy, $(\zeta^-\partial_-+\zeta^+\partial_+)\tilde{\theta}_+=0$, where $\zeta^+\zeta^->0$. Moreover, using conditions (\ref{HWormhole}) and the result (\ref{ConfTheta}), we find that to have a Hayward wormhole in the new frame, the following conditions need to be satisfied,
\begin{equation}\label{ConfHWormhole}
\begin{gathered}
\theta_{\pm}+2\partial_\pm\Omega\big|_S=0,\qquad
\partial_\pm\theta_{\pm}+2\partial_\pm\partial_\pm\Omega\big|_S>0,\\
\partial_\mp\theta_{\pm}+2\partial_\mp\partial_\pm\Omega\big|_S<0.\\
\end{gathered}
\end{equation}
As for the temporal character of the hypersurface, we easily see that if a Hayward temporal wormhole exists in the old frame another temporal one might exist in the new frame regardless of whether the conformal factor satisfies along the timelike hypersurface, $(\zeta^-\partial_-+\zeta^+\partial_+)\partial_+\Omega=0$. In addition, the equality in (\ref{ConfHWormhole}) imposes on $\Omega$ to be independent of $u^\pm$. The inequality simply constrains the degree of convergence of light rays near the throat. On the other hand, if no such wormhole exists in the original frame, that is, if the original expansion $\theta_\pm$ violates either one or all the constraints (\ref{HWormhole}), a temporal Hayward wormhole might still exist if the conformal factor $\Omega$ is chosen such that conditions (\ref{ConfHWormhole}) are satisfied. 

Yet, another interesting possibility is the following. As we mentioned it above, a Hayward wormhole is necessarily a Hochberg-Visser wormhole but the converse is not true. In the conformal frame, however, one might get a purely Hochberg-Visser wormhole by starting from a Hayward wormhole. In fact, if the conformal factor is chosen such that $(\zeta^-\partial_-+\zeta^+\partial_+)(\theta_++\partial_+\Omega)=0$ for a spacelike vector, $\zeta^+\zeta^-<0$, and such that the last inequality in (\ref{ConfHWormhole}) is not satisfied but the second inequality is, then a pure Hochberg-Visser wormhole arises instead.

We now examine the Maeda-Harada-Carr wormhole. In this case we have the additional arbitrary spacelike radial vector $\zeta^\mu$. Whatever scaling form such a vector takes under a conformal transformation, though, both the equality and the inequality in (\ref{MHCWormhole}) will not be affected as only an overall positive factor would be introduced. On the other hand, from the metric form (\ref{MetricForCosmologicalWormholes}) we have $\tilde{R}=e^{\Omega}R$ and we shall use the following transformation of the two-dimensional Christoffel symbols $\tilde{\Gamma}_{AB}^C=\Gamma_{AB}^C+\delta_A^C\Omega_{,B}+\delta_B^C\Omega_{,A}-g_{AB}\Omega^{,C}$. Thus, to have such a wormhole in the new frame one needs the following two conditions to be simultaneously satisfied on the 2-surface that is supposed to represent the wormhole's throat,
\begin{align}\label{ConfMHCWormhole}
&\zeta^A\left(R_{|A}+R\Omega_{|A}\right)\Big|_S=0\nonumber\\
&\zeta^A\zeta^B\left(R_{|AB}+R\Omega_{|AB}-R\Omega_{|A}\Omega_{|B}\right)\nonumber\\
&\qquad+\zeta^A\zeta^Bg_{AB}\left(R_{|C}\Omega^{|C}+R\Omega_{|C}\Omega^{|C}\right)\Big|_S>0.
\end{align}
It becomes clear here again that one can always use a conformal factor such that both the equality and the inequality become satisfied and have a Maeda-Harada-Carr cosmological wormhole even if none exists in the original frame, {\textit{i.e.}}, even if conditions (\ref{MHCWormhole}) are not satisfied. On the other, it could easily happen that conditions (\ref{MHCWormhole}) be satisfied but no conformal factor could be found to make conditions (\ref{ConfMHCWormhole}) satisfied. In this case, such a wormhole becomes inexistent in the new frame. 

We finally end this section by examining the conformal transformation of a Tomikawa-Izumi-Shiromizu wormhole. Using definition (\ref{TISWormhole}), we easily find that to have such a wormhole in the conformal frame one needs the following conditions to be simultaneously satisfied on the 2-surface that is supposed to represent the wormhole's throat,
\begin{equation}
\begin{gathered}\label{ConfTISWormhole}
\theta_+-\theta_-+2\left(\partial_+\Omega-\partial_-\Omega\right)\Big|_S=0,\\
(\partial_+-\partial_-)\left[\theta_+-\theta_-+2\left(\partial_+\Omega-\partial_-\Omega\right)\right]\Big|_S>0.
\end{gathered}
\end{equation}
In writing the inequality, we have used the observation, also made below Eq.~(\ref{NonInvariance}), that the conformal transformation would only introduce the overall positive multiplicative factor $e^\Omega$ on the operator $\partial_+-\partial_-$ when going from the non-affine parametrization $\tilde{u}^\pm$ to the affine parametrization $u^\pm$.

From these two conditions, the same scenario that appeared for the previous wormhole definitions repeats itself; namely, a Tomikawa-Izumi-Shiromizu wormhole could arise in the conformal frame even if in the original frame there was none, provided that the conformal factor is chosen to satisfy both conditions (\ref{ConfTISWormhole}). Conversely, the existence of such a wormhole in the original frame does not guarantee its existence in the new frame.
\section{Worked out Examples}\label{SecIV}
In this section we are going to pick up a simple metric and investigate the behavior of the various definitions under Weyl's conformal transformation. The metric we choose should be simple enough that for each definition it constitutes an easy example for illustrating all the key features pointed out about the conformal transformation of that specific definition.
\subsubsection{The simple prescription}
We are going to start by providing a worked out example for the simple prescription (\ref{Prescription}). For that purpose, we shall use a simple static wormhole metric for which it is easy to find the solution to the equation, $R_{,r}=0$, involving the derivative of the areal radius $R$. In fact, the following Brans class IV metric \cite{Brans},
\begin{equation}\label{BransIV}
{\rm{d}}s^2=-e^{-2\alpha/r}{\rm{d}}t^2+e^{2\beta/r}({\rm{d}}r^2+r^2{\rm{d}}\Theta^2).
\end{equation}
written in isotropic coordinates and in which $r>0$, will nicely fit into this purpose. The constant $\alpha$ is arbitrary while the constant $\beta$ is a function of $\alpha$ and the Brans-Dicke parameter $\omega$ of Brans-Dicke theory \cite{BransDicke}, but whose exact expression need not concern us here\footnote{For completeness, we give here the full relation between $\alpha$, $\beta$, and $\omega$ as given in Brans' paper \cite{Brans}: $\beta=\alpha\left(\frac{\omega+1\pm\sqrt{-(2\omega+3)}}{\omega+2}\right)$.}. Such a metric has been studied in Ref.~\cite{FFS} by relying only on the prescription (\ref{Prescription}). The areal radius can be read off directly from metric (\ref{BransIV}): $R=re^{\beta/r}$. So, condition (\ref{StaticDR/Dl}), extracted from the simple prescription (\ref{Prescription}), yields the following equation in $r$ \cite{FFS},
\begin{equation}\label{BransCondition}
\frac{\left(r-\beta\right)^2}{r^2}=0.
\end{equation}
The double-root of this equation, $r_*=\beta$, should correspond therefore to a wormhole throat. We assume that $\alpha$ is chosen such that $\beta>0$, so that $r_*>0$. Notice that in this simple prescription, there is no indication to assure us that the radius $R$ really reached its minimum value as no second derivative is involved in the prescription. All one can say in addition is that there should not be any horizon, as there is no single-root in this case.

By conformally transforming now the metric (\ref{BransIV}) using some radial- and time-dependent function  $\Omega(t,r)$, chosen as such in order to preserve the spherical symmetry of the spacetime, the metric (\ref{BransIV}) becomes ${\rm{d}}\tilde{s}^2=e^{2\Omega(t,r)}{\rm{d}}s^2$ and condition (\ref{NewCondition}) gives the following equation:
\begin{equation}\label{ConfBransCondition}
\left(r^2\Omega_{,r}+r-\beta\right)^2-r^4e^{2(\alpha+\beta)/r}\Omega_{,t}^2=0.
\end{equation}

Let us begin with a time-independent conformal factor. In this case, $\Omega_{,t}=0$. Then, by choosing, for simplicity, $\Omega(r)=\beta/r$, which is regular everywhere except at the origin, Eq.~(\ref{ConfBransCondition}) yields the double-root $r_*=2\beta$. Since in this case there is no single-root we should conclude that there is no horizon, but just a wormhole whose throat is sitting at the radial coordinate $r_*$. 

By using a time-dependent conformal factor, however, one can choose a function $\Omega(t,r)$ such that Eq.~(\ref{ConfBransCondition}) admits just a single-root, but no double-roots. In fact, such a function exists and it can be chosen to be smooth and regular for all time $t$. One simple choice consists of,
$\Omega(t)=\tanh(t/4\beta)$. Substituting this function inside Eq.~(\ref{ConfBransCondition}) yields different single-roots for different values of the coordinate time $t$. One example of such a single-root is again at $r_0=2\beta$. It occurs at the specific coordinate time $t_0=4\beta\coth^{-1}[e^{-(\alpha+\beta)/2\beta}]$. Of course, here we assume that the Brans-Dicke parameter $\omega$, and hence the constants $\alpha$ and $\beta$, are such that the argument $e^{-(\alpha+\beta)/2\beta}$ is greater in absolute value than, or equals, 1. As time goes by, the location $r_0$ of the horizon changes. Thus, we have found a conformal transformation for which the simple prescription (\ref{Prescription}) makes a black hole horizon arise in the new frame even though the original frame did not contain any horizon. Such a transformation had to be time-dependent as a static one does not allow it.

Another noticeable location that could host a wormhole throat in the conformal frame within this definition, but that could not within Hochberg-Visser and Hayward definitions, as we shall see below, is $r_0=\beta$. Indeed, with the function $\Omega(t,r)=\beta/r+\tanh(t/\beta)$, the time coordinate of such a throat is then $t_0=\beta\coth^{-1}[e^{-2(\alpha+\beta)/\beta}]$.
\subsubsection{Black hole horizon}
In this subsection, we use the same metric (\ref{BransIV}) that we used to illustrate the consequences of relying on the simple prescription (\ref{Prescription}) for detecting apparent horizons. Here we focus on the "sophisticated" definition of the black hole horizon to investigate its presence and its conformal behavior within such a metric. For that purpose we start by giving two null vectors for such a geometry:
\begin{equation}\label{BransNullVectors}
l_\pm^\mu=\frac{1}{\sqrt{2}}\left(e^{\alpha/r},\pm e^{-\beta/r},0,0\right).
\end{equation}
Notice that a direct computation reveals that these vectors obey a non-affine geodesic equation. As remarked in subsection \ref{Definitions}, however, this will not affect our subsequent analysis of the extremality and minimality conditions in the original frame. By using the definition $h^{\mu\nu}\nabla_\mu l_{\pm\nu}$ of the null congruence's expansions, and the transverse metric $h_{\mu\nu}=e^{2\beta/r}\times\text{diag}(0,0,r^2,r^2\sin^2\vartheta)$, we easily compute the expansions $\theta_\pm$ and the variation of the outgoing expansion $\partial_\mp
\theta_+=l_\mp^\mu\partial_\mu
\theta_+$ along the ingoing (outgoing) null geodesics, respectively. We find,
\begin{equation}\label{BransExpansions}
\begin{gathered}
\theta_\pm=\pm\frac{\sqrt2e^{-\beta/r}}{r^2}(r-\beta),\\
\partial_\mp\theta_+=\pm\frac{e^{-2\beta/r}}{r^4}(r^2-3\beta r+\beta^2).
\end{gathered}
\end{equation}
From these expressions we easily check that the first and last conditions in (\ref{BHHorizon}) are satisfied at $r_0=\beta$, but that the second condition is violated there. Therefore, we conclude that there cannot be any black hole horizon in this geometry. Notice that we can also understand from expressions (\ref{BransExpansions}) why we cannot have a black hole horizon. In fact, the first identity in (\ref{BransExpansions}) implies that $\theta_+\theta_-<0$ for all $r\neq0$. This indicates that no trapped surfaces exist in such a geometry as every surface is untrapped.

We now perform again a Weyl conformal transformation on the metric (\ref{BransIV}) using a time-dependent conformal exponent $\Omega(t,r)$ that might be chosen to be radial to preserve the spherical symmetry for simplicity. Using expressions (\ref{BransExpansions}), we obtain, by relying on the requirements (\ref{ConfBHHorizon}), the conditions for having a black hole horizon in the new frame as follows:
\begin{align}\label{ConfBransBH}
&r^2\Omega_{,r}+r-\beta+ r^2e^{(\alpha+\beta)/r}\Omega_{,t}\big|_{H}=0,\nonumber\\
&r^2\Omega,r+r-\beta-r^2e^{(\alpha+\beta)/r}\Omega_{,t}\big|_{H}>0,\nonumber\\
&r^2-3\beta r+\beta^2-r^4\Omega_{,rr}-\beta r^2\Omega_{,r}\nonumber\\
&\qquad\qquad\qquad+r^4e^{2(\alpha+\beta)/r}\Omega_{,tt}+\alpha r^2e^{(\alpha+\beta)/r}\Omega_{,t}\Big|_H<0.
\end{align}
In writing the first inequality we have multiplied both sides by $-1$. Comparing the equality and the first inequality with equation (\ref{ConfBransCondition}) we can already see the precision added by the definition (\ref{BHHorizon}) for black hole horizons over the simple prescription (\ref{Prescription}) for detecting them. Indeed, the left-hand side of equation (\ref{BHHorizon}) is just the product of the left-hand sides of the equation and the first inequality in (\ref{ConfBransBH}). So, while the simple prescription requires only the vanishing of a product of two spacetime-dependent factors, the rigorous definition requires the vanishing of just one factor of such a product and constrains the other factor to be positive. Moreover, a third inequality is required that involves the second derivatives of the conformal exponent that drives the variation of the null congruences' expansions in the new frame.

Plugging inside conditions (\ref{ConfBransBH}) the simple function $\Omega(r)=\beta/r$ we tested on equation (\ref{ConfBransCondition}), we easily check that the first inequality is violated at $r_0=\beta$ for which the equation is satisfied. This confirms that $r_0=\beta$ cannot be the location of a black hole in the conformal frame obtained with a time-independent conformal factor. Moreover, just by comparing the equation and the inequality, after putting $\Omega_{,t}=0$, we clearly see that no black hole horizon could ever arise from a static conformal transformation. This being the case only for this spherically symmetric spacetime though.

Plugging now inside (\ref{ConfBransBH}) the time-dependent function $\Omega(t)=\tanh(t/4\beta)$ we see that the location $r_0=2\beta$ that we previously found hosting an apparent horizon at time $t_0=4\beta\coth^{-1}[e^{-(\alpha+\beta)/2\beta}]$ does not host one here. Instead, we find that $r_0=2\beta$ can be the location of a black hole horizon at that time $t_0$ if one uses the function $\Omega(t)=-\tanh(t/4\beta)$; that is, having a contracting universe instead of an expanding one. This required time-reversal implies, as one can easily verify, that by substituting $\theta_-$ for $\theta_+$ and $\partial_-$ for $\partial_+$ in the definition (\ref{BHHorizon}) conditions (\ref{ConfBransBH}) will indeed be satisfied at that $r_0$ and at that $t_0$ within an expanding universe. Therefore, what was thought would be a black hole horizon in the conformal frame at $r_0$ based on the simple prescription will actually be a white hole according to the rigorous definition. 

\subsubsection{Hochberg-Visser wormhole}
Since the example chosen by the authors in Ref.~\cite{HochbergVisser} to illustrate their wormhole definition was already the {\textit{conformal}} Morris-Thorne wormhole metric \cite{MorrisThorn}, we are not going to repeat that analysis in this subsection. Suffice it to mention here that after computing explicitly the expansions $\theta_\pm$ using the Morris-Thorne metric in the original frame and then substituting in our conditions (\ref{ConfHVWormhole}) one gets exactly the same conditions obtained in Ref.~\cite{HochbergVisser} by starting right away with the conformal frame.

In order to illustrate our results concerning Hochberg-Visser's definition we prefer, therefore, to use again the same Brans class IV metric already worked out in detail above. In fact, the only additional expression that is needed here, besides the expansions $\theta_\pm$ given by expressions (\ref{BransExpansions}), is the derivative $\partial_+\theta_+$ from the second identity there. It is then straightforward to check that at $r_0=\beta$ we have indeed a Hochberg-Visser wormhole as the two conditions $\theta_+|_S=0$ and $\partial_+\theta_+|_S>0$ are satisfied there\footnote{For definiteness, we shall only work here with one of the two available sets of conditions (\ref{HVWormhole}), choosing the set with the plus signs.}.

Moving now to the conformal frame with an exponent $\Omega(t)$ depending only on time for simplicity, we easily conclude from conditions (\ref{ConfHVWormhole}) that to have such a wormhole again in that frame, we need the following requirements to be satisfied on the 2-surface:
\begin{align}\label{ConfBransHVWormhole}
&r-\beta+ r^2e^{(\alpha+\beta)/r}\Omega_{,t}\big|_{S}=0,\nonumber\\
&r^2-3\beta r+\beta^2-r^4e^{2(\alpha+\beta)/r}\Omega_{,tt}+\alpha r^2e^{(\alpha+\beta)/r}\Omega_{,t}\Big|_S<0.
\end{align}
In writing the inequality in (\ref{ConfBransHVWormhole}) we have multiplied both sides by $-1$. We see that for $\Omega(t)=\tanh(t/4\beta)$, the location $r_0=\beta$ cannot host a Hochberg-Visser wormhole at any time and neither can the location $r_0=2\beta$, at which a horizon is supposed to appear. More importantly, in contrast to equation (\ref{ConfBransCondition}), which would always allow for a wormhole at $r<\beta$ or at $r>\beta$ in an expanding universe ($\Omega_{,t}>0$), conditions (\ref{ConfBransHVWormhole}) allow for a wormhole at $r<\beta$ only in an expanding universe and at $r>\beta$ only in a contracting universe.

\subsubsection{Hayward wormhole}
As we explained it above, every Hayward wormhole is necessarily a Hochberg-Visser wormhole as well, but the converse is not true. We also argued that it is possible, using a Weyl conformal transformation, to turn a Hayward wormhole into a purely Hochberg-Visser one. Our aim in this subsection is to illustrate that conclusion. For that purpose, we use again Brans class IV metric (\ref{BransIV}). Then, having at our disposal the necessary components (\ref{BransExpansions}), we can ascertain that the wormhole, with its throat sitting at $r_0=\beta$, is also a Hayward wormhole as expressions (\ref{BransExpansions}) make all three conditions (\ref{HWormhole}) satisfied there.

Moving to the conformal frame with $\Omega=\Omega(t)$ for simplicity, we easily get from the requirements (\ref{ConfHWormhole}) and expressions (\ref{BransExpansions}) the explicit conditions to be satisfied in order to still have the Hayward wormhole in the new frame. They read,
\begin{align}\label{ConfBransHWormhole}
&r-\beta+ r^2e^{(\alpha+\beta)/r}\Omega_{,t}\big|_{S}=0,\nonumber\\
&r^2-3\beta r+\beta^2-r^4e^{2(\alpha+\beta)/r}\Omega_{,tt}+\alpha r^2e^{(\alpha+\beta)/r}\Omega_{,t}\Big|_S<0,\nonumber\\
&r^2-3\beta r+\beta^2+r^4e^{2(\alpha+\beta)/r}\Omega_{,tt}+\alpha r^2e^{(\alpha+\beta)/r}\Omega_{,t}\Big|_S<0.
\end{align}

Just as for Hochberg-Visser wormhole, we see that for $\Omega(t)=\tanh(t/4\beta)$, the location $r_0=\beta$ cannot host a Hayward wormhole at any time and neither can the location $r_0=2\beta$, at which a horizon is supposed to appear. More importantly, however, we see that it might happen that, among the conditions (\ref{ConfBransHWormhole}), the equation and the inequality in the middle are satisfied but that the last inequality is violated. In fact, for $\Omega(t)=\tanh(2t/\beta)$ the location $r_0=\beta/2$ corresponding to the coordinate time $t_0=\frac{\beta}{2}\coth^{-1}[e^{-2(\alpha+\beta)/\beta}]$ is an example. This means that we have obtained in this case a purely Hochberg-Visser wormhole starting from a Hayward wormhole, albeit at a single instant of time.  
\subsubsection{Maeda-Harada-Carr wormhole}
The Maeda-Harada-Carr definition for cosmological wormholes has been illustrated by the authors themselves in Ref.~\cite{MHC} by using, among other things, the cosmological Ellis wormhole. Such a metric is very interesting from the point of view of conformal transformations as it is conformal to the static Ellis wormhole. However, such a metric has also been the subject of a detailed illustration in Ref.~\cite{TIS} for the Tomikawa-Izumi-Shiromizu definition. Furthermore, it is not hard to recover the formulas found there by using our conformally transformed conditions (\ref{ConfMHCWormhole}) instead. Therefore, we are going to use here again and in the next subsection Brans class IV solution (\ref{BransIV}). This will allow us to confront these last two definitions of wormholes in the light of the conformal transformation.

Let us then start by rewriting the metric (\ref{BransIV}) in the double-null coordinates system. To achieve that, we perform the following successive coordinate redefinitions,
\begin{equation}\label{TowardsNullCoordinates}
{\rm{d}}\rho=e^{(\alpha+\beta)/r}{\rm{d}}r,\qquad u=\frac{t-\rho}{\sqrt2},\qquad v=\frac{t+\rho}{\sqrt2}.
\end{equation}
The metric (\ref{BransIV}) then takes on the standard form (\ref{MetricForCosmologicalWormholes}) with,
$f(u,v)=-\alpha/r(u,v)$ and $R(u,v)=r(u,v)e^{\beta/r(u,v)}$. For a radial spacelike vector $\zeta^\mu$ (such that $\zeta^u\zeta^v<0$), and using the non-vanishing Christoffel symbols for the two-dimensional space, $\Gamma_{uu}^u=2f_{,u}$ and $\Gamma_{vv}^v=2f_{,v}$, as well as the chain rule, 
\begin{equation}
\partial_{u,v}=\frac{1}{\sqrt2}\left(\partial_t\mp e^{-(\alpha+\beta)/r}\partial_r\right),
\end{equation}
for the null coordinates $u$ and $v$, respectively, the conditions (\ref{MHCWormhole}) yield the following two requirements to be satisfied on the 2-surface:
\begin{equation}\label{HMCOnBrans}
(\zeta^v-\zeta^u)R_{,r}\big|_S=0,\qquad(\zeta^v-\zeta^u)^2R_{,rr}\big|_S>0.
\end{equation}
To arrive at the inequality, we have taken into account the fact that the equation implies $R_{,r}\big|_S=0$. Now, these conditions are just expressing the usual extremality and minimality of the areal radius $R$ on a spacelike hypersurface. By substituting $R=re^{\beta/r}$, it is easy to check that the equation implies $r_0=\beta$ while the inequality is equivalent to $\beta e^{\beta/r}/r^3>0$, which is always satisfied. Therefore, we see in this case that, as for the previous two definitions, the location $r_0=\beta$ constitutes a Maeda-Harada-Carr wormhole as well.

When going to the conformal frame, it is sufficient to take the conformal exponent $\Omega=\Omega(t)$ depending only on time. In fact, besides simplifying the formulas, this choice is already capable, thanks to the time-derivatives of $\Omega$, of displaying the distinctive feature of this definition which consists in separating the contribution weighted by the difference $(\zeta^v-\zeta^u)$ from the contribution weighted by the sum $(\zeta^v+\zeta^u)$. The spatial radial vector acquires thus an important role in this definition. Indeed, the conditions (\ref{ConfMHCWormhole}) yield, after substituting $R=re^{\beta/r}$, the following:
\begin{align}\label{ConfBransHMCWormhole}
&(\zeta^v-\zeta^u)(r-\beta)+(\zeta^v+\zeta^u)r^2e^{(\alpha+\beta)/r}\Omega_{,t}\Big|_S=0,\nonumber\\
&(\zeta^v-\zeta^u)^2\left[(\alpha+\beta)r-\alpha\beta-r^4e^{2(\alpha+\beta)/r}\Omega_{,t}^2\right]\nonumber\\
&+(\zeta^v+\zeta^u)^2r^4e^{2(\alpha+\beta)/r}\Omega_{,tt}-2\alpha({\zeta^v}^2+{\zeta^u}^2)\nonumber\\
&\qquad\qquad\qquad\qquad\qquad\times\left[r-\beta+r^2e^{(\alpha+\beta)/r}\Omega_{,t}\right]\bigg|_S>0.
\end{align}
From these conditions we clearly see the difference with respect to (\ref{ConfBransHVWormhole}) and (\ref{ConfBransHWormhole}) brought by the wormhole definition (\ref{MHCWormhole}). The radial spatial, but most importantly arbitrary, vector $\zeta^\mu$ plays a more decisive role in the conformal frame than it does in the original frame. To fully appreciate such difference we start by checking the -- by now familiar -- locations $r_0=\beta$ and $r_0=2\beta$. 

In fact, we immediately check that, in contrast to the Hochberg-Visser and Hayward definitions, for which the location $r_0=\beta$ can no longer host a wormhole throat in the conformal frame, according to the Maeda-Harada-Carr definition a wormhole throat can appear at such location in the conformal frame. Indeed, the equation in (\ref{ConfBransHMCWormhole}) is satisfied at $r_0=\beta$ for all $t$ provided that one chooses the spatial radial vector $\zeta^\mu$ such that $\zeta^v=-\zeta^u$. Then, the inequality holds for all $t$ provided only that the conformal exponent $\Omega(t)$ satisfies,
\begin{equation}
\left[\beta^2e^{2(\alpha+\beta)/\beta}+\alpha e^{(\alpha+\beta)/\beta}\right]\Omega_{,t}<1.
\end{equation}

Next, by choosing, for definiteness, $\zeta^v=-3\zeta^u$ we also easily check that, in contrast to what happens when using either the Hochberg-Visser or the Hayward wormhole definitions, the location $r_0=2\beta$ can host a Maeda-Harada-Carr wormhole throat. Indeed, with such a choice for the spatial vector $\zeta^\mu$, the function $\Omega(t)=-\tanh(t/2\beta)$ giving a contracting universe would solve the equation in (\ref{ConfBransHMCWormhole}) at $r_0=2\beta$ at the time coordinate $t_0=2\beta\coth^{-1}[e^{-(\alpha+\beta)/2\beta}]$. To satisfy the inequality in (\ref{ConfBransHMCWormhole}) one then merely needs to choose $\alpha$ and $\beta$ such that $3\alpha\beta>4\beta^2$.

Thus the arbitrariness of the spatial vector $\zeta^\mu$ involved in the definition of the Maeda-Harada-Carr wormhole makes the latter depart from the previous definitions in a more accentuated manner especially under a conformal transformation. 

\subsubsection{Tomikawa-Izumi-Shiromizu wormhole}
To illustrate this last definition of wormholes we use again Brans metric (\ref{BransIV}). Recall that the definition (\ref{TISWormhole}) requires the use of an affine parametrization on the null geodesics. Now, however, the null vectors as given by expressions (\ref{BransNullVectors}) do not obey an affine geodesic equation as can easily be checked. As we saw above, a re-parametrization is then required in order to still be able to use $\partial_\pm=l_\pm^{\mu}\partial_{\pm\mu}$ in our calculations when working with the definition (\ref{TISWormhole}). The needed re-parametrization is of the form given in the transformation (\ref{ParametrizationChange}). Since this introduces only a positive multiplicative factor, the two conditions (\ref{TISWormhole}) taken together will then remain unaffected when using a non-affine parameterization. In fact, the condition imposed by the equation there always cancels the additional terms that would come from the derivatives of such a multiplicative factor even within the inequality. 

Therefore, substituting inside conditions (\ref{TISWormhole}) the corresponding congruence expansions (\ref{BransExpansions}), the conditions that would make the metric (\ref{BransIV}) a Tomikawa-Izumi-Shiromizu wormhole read,
\begin{equation}\label{BransTISWormhole}
\begin{gathered}
r-\beta\big|_S=0\qquad \text{and}\qquad r^2-3\beta r+\beta^2\big|_S<0.
\end{gathered}
\end{equation}
Again, we see that such conditions both hold on the throat $r_0=\beta$, thus making the metric (\ref{BransIV}) also a Tomikawa-Izumi-Shiromizu wormhole. 

Going to the conformal frame, using a conformal exponent $\Omega=\Omega(t,r)$ that depends both on time and the radial coordinate $r$, conditions (\ref{ConfTISWormhole}) yield the following two requirements to be satisfied on the 2-surface $S$ in order to have a Tomikawa-Izumi-Shiromizu wormhole in the new frame,
\begin{equation}\label{ConfBransTISWormhole}
\begin{gathered}
r-\beta+r^2\Omega_{,r}\big|_S=0,\\
r^2-3\beta r+\beta^2-r^4\Omega_{,rr}-\beta r^2\Omega_{,r}\big|_S<0.
\end{gathered}
\end{equation}

We first notice in these conditions the peculiar feature, not shared by any of the other definitions under a Weyl transformation, which is the absence of any time-derivative of the conformal exponent $\Omega(t,r)$. This is hardly surprising though as
the definition should remain time-symmetric in the conformal frame for it only involves the difference between the ingoing and outgoing expansions as well as the spacelike derivative of this difference.

An example of the difference with respect to the Maeda-Harada-Carr definition is that, in contrast to the latter, the location $r_0=\beta$ can never host a Tomikawa-Izumi-Shiromizu wormhole throat in the conformal frame, whereas the location $r_0=2\beta$ can. The conformal exponent that allows that being $\Omega(r)=\beta/r$.   

The left-hand sides of both the equality and the inequality in conditions (\ref{ConfBransTISWormhole}) are actually what would appear in the left-hand sides of the conditions (\ref{ConfBransHVWormhole}) and (\ref{ConfBransHWormhole}) for the Hochberg-Visser and Hayward definitions, respectively, had we written the latter for a radial-dependent exponent factor $\Omega(r)$. This indeed can be expected just by inspecting the conformal black hole conditions we wrote using the more general function $\Omega(t,r)$. Thus, we see that, under the Weyl conformal transformation, the Tomikawa-Izumi-Shiromizu definition behaves like a truncated version of the previous definitions, being insensible to any time dependence of the conformal factor that distorts spacetime to create the new frame.   
\section{Conclusion \& discussion}
We have examined the behavior under the Weyl conformal transformation of various definitions proposed in the literature for wormholes and black holes' horizon. We then illustrated the general results found using worked out examples based on a Brans class IV solution of Brans-Dicke theory, because it happens that, not only such a metric contains a non-trivial wormhole, but it is also simple enough to illustrate each one of the various definitions examined in this paper.

From this study it became clear why things look different from the perspective of one frame or the other despite the usual argument for the physical equivalence of the two frames. The reason is that a Weyl conformal transformation does change spacetime geometry, and horizons, as well as wormholes, are nothing but geometry. We saw that almost everything is possible under this kind of conformal transformation of spacetime and that black holes or wormholes are not forbidden from appearing in the new frame even if they do not exist in the old frame. Wormholes might disappear or be transformed from one type to another, and black holes might suddenly disappear or be created. We saw, however, that subtle differences exist between the simple prescription usually used in the literature for locating wormholes and apparent horizons in spherically symmetric spacetimes and the more rigorous definitions. We found that these differences manifest themselves more noticeably under the conformal transformation.

It is easy to see the reason behind the conformal behavior of black holes' horizon and wormholes from the way Einstein equations transform. Under a Weyl transformation, equations $G_{\mu\nu}=8\pi GT_{\mu\nu}$ transform into $\tilde{G}_{\mu\nu}=8\pi GT_{\mu\nu}+T^{\Omega}_{\mu\nu}$, where $T^{\Omega}_{\mu\nu}$ is an $\Omega$-induced energy-momentum tensor. Thus, a given solution of the original Einstein equations gets necessarily mapped under the Weyl transformation into a different solution with a different effective source. Although the study conducted here is independent of the dynamics of spacetime, as no field equation was involved, one might still wonder if this behavior would be preserved for black holes and wormholes within modified gravity theories and, if so, how could it be understood from the point of view of the dynamics. The answer to such a question can easily be obtained by referring to the very recent paper {\cite{Karam}} in which the field equations of scalar-tensor theories of gravity have been written using frame-invariant quantities that depend only on the scalar-tensor gravity model. For a theory with the action $\int{\rm d}^4x\sqrt{-g}\left[\frac{1}{2}A(\phi)R-\frac{1}{2}B(\phi)(\nabla\phi)^2-V(\phi)\right]+S_m(e^{2\sigma(\phi)}g_{\mu\nu},\psi)$, where $A(\phi)$, $B(\phi)$, $V(\phi)$ and $\sigma(\phi)$ are arbitrary regular functions of the scalar field $\phi$, and $S_m$ is the action of the matter fields $\psi$, such frame-invariant field equations read \cite{Karam}, $\hat{G}_{\mu\nu}=16\pi G\hat{T}_{\mu\nu}+2\hat{\nabla}_\mu\mathcal{I}_3\hat{\nabla}_\nu\mathcal{I}_3+\hat{g}_{\mu\nu}\left(\hat{\nabla}_{\rho}\mathcal{I}_{3}\hat{\nabla}^{\rho}\mathcal{I}_{3}+\mathcal{I}_{2}\right)$. In these equations the hatted quantities are built from the frame-invariant metric $\hat{g}_{\mu\nu}=A(\phi)g_{\mu\nu}$, and the frame-invariants $\mathcal{I}_2$ and $\mathcal{I}_3$ are built from the functions $A$, $B$, $V$ and $\sigma$. This shows that the field equations can indeed be put in a form that is invariant under conformal transformations and this might then suggest the existence of a frame-invariant solution, in contrast to our conclusion above. However, as the frame-invariant Einstein tensor $\hat{G}_{\mu\nu}$ is built from the metric $\hat{g}_{\mu\nu}$ instead of the physical metric $g_{\mu\nu}$, the solution to such frame-invariant field equations would not represent a real physical solution. This observation can actually be used the other way around. In fact, we can now argue that, since the only field equations that are frame-invariant require the use of a non-physical metric, no physical solution can fully be preserved under a Weyl transformation in such modified gravity theories either.

These results might hint at a possible non-trivial physical meaning of the Weyl conformal transformation and might well point against the usual argument according to which a rescaling of our physical units will always cancel locally the physical effect that such a transformation of spacetime would produce. However, we believe that the issue of the physical equivalence of the two frames needs a more specialized investigation which will be attempted in more depth in future works.

As a final remark we would like to mention here that by having a black hole arise in the conformal frame even if it did not exist in the original frame might suggest a possible violation of the no-hair theorem as the conformal exponent $\Omega(t,r)$ would play the role of a scalar field in the transformed Einstein equations. As we saw in our worked out examples, however, for a black hole horizon to arise the would-be scalar field $\Omega(t,r)$ should depend on time, thus invalidating the staticity assumption that gave rise to the no-hair theorem \cite{Hawking}. 

\begin{acknowledgments}
The author greatly benefited from the enthusiasm of Valerio Faraoni towards the author's "artistic" illustrations and his encouragements to include them, and would like to thank the anonymous referee for pointing out Ref.~\cite{Karam}.

This work is supported by the Natural Sciences \& Engineering Research Council of Canada (NSERC Grant No.~2017-05388).
\end{acknowledgments}

\end{document}